\documentclass{article}
\usepackage{spconf,amsmath,graphicx, comment, color}

\title{Rhythm-Flexible Voice Conversion without Parallel Data Using Cycle-GAN over Phoneme Posteriorgram Sequences}
\name{Cheng-chieh Yeh$^1$, Po-chun Hsu$^1$, Ju-chieh Chou$^1$, Hung-yi Lee$^1$, Lin-shan Lee$^1$}
\address{
  $^1$College of Electrical Engineering and Computer Science, National Taiwan University\\
  \{r06942067, b03901071, r06922020, hungyilee\}@ntu.edu.tw, lslee@gate.sinica.edu.tw
}
\begin{document}
%
\maketitle
\begin{abstract}
Speaking rate refers to the average number of phonemes within some unit time, while the rhythmic patterns refer to duration distributions for realizations of different phonemes within different phonetic structures.
Both are key components of prosody in speech, which is different for different speakers.
Models like cycle-consistent adversarial network (Cycle-GAN) and variational auto-encoder (VAE) have been successfully applied to voice conversion tasks without parallel data.
However, due to the neural network architectures and feature vectors chosen for these approaches, the length of the predicted utterance has to be fixed to that of the input utterance, which limits the flexibility in mimicking the speaking rates and rhythmic patterns for the target speaker.
On the other hand, sequence-to-sequence learning model was used to remove the above length constraint, but parallel training data are needed.
In this paper, we propose an approach utilizing sequence-to-sequence model trained with unsupervised Cycle-GAN to perform the transformation between the phoneme posteriorgram sequences for different speakers.
In this way, the length constraint mentioned above is removed to offer rhythm-flexible voice conversion without requiring parallel data. Preliminary evaluation on two datasets showed very encouraging results.
\end{abstract}


\section{Introduction}
\label{sec:intro}
Voice conversion (VC) is a task aiming to convert the speech signals from a certain acoustic domain to another while keeping the linguistic content the same.
Examples of considered acoustic domains include not only the speaker identity, but many other factors orthogonal to the linguistic content such as speaking style, speaking rate~\cite{rentzos2003transformation}, noise condition, emotion~\cite{aihara2012gmm, kawanami2003gmm}, accent~\cite{oyamada2017non}, etc., with potential applications ranging from speech enhancement~\cite{kain2007improving, rudzicz2011acoustic}, computer-assisted pronunciation training for non-native language learner~\cite{oyamada2017non}, speaking assistance~\cite{nakamura2012speaking}, to speaker identity conversion~\cite{stylianou1998continuous, kain1998spectral, saito2011one, kinnunen2017non}, to name a few.

When speaker identity conversion is considered, in addition to the fact that the same phoneme sounds different when produced by different speakers, it is well known that the prosody can also be very different for different speakers.
The prosody of speech includes not only the pitch range, but at least the speaking rate and the rhythmic patterns. 
While speaking rate refers primarily to the average number of phonemes produced within some unit time, the rhythmic pattern refers to the duration distributions for realizations of different phonemes within different phonetic structures.
It is obvious that the speaking rate and rhythmic patterns are very different for different speakers.
When the goal is to mimic the voice characteristics of a specific speaker, it is important that the prosody including the speaking rate and the rhythmic patterns of the target speaker is reproduced.
This is why flexible speaking rates and rhythmic patterns are highly desired for voice conversion (VC).

The many approaches proposed for VC may be in most cases classified into two types: text-independent and text-dependent.
Text-independent VC directly predicts the target speech signals based on the source speech signals without considering the linguistic content or text.
This is usually achieved with acoustic models such as Gaussian mixture models (GMMs)~\cite{stylianou1998continuous, toda2007voice} or deep neural networks (DNNs)~\cite{desai2009voice}.
Text-dependent VC, on the other hand, converts speech signals through the textual information.
That is, a speech recognizer is used to estimate the textual information from the source speech and a speech synthesizer is used to predict the target speech from the textual information.
The conversion units for text-dependent VC are usually rougher (e.g., phonemes, characters or words) than those used in text-independent VC (e.g., frames).
Approaches recently proposed using phoneme posteriorgram vectors as the conversion unit~\cite{sun2016personalized, xie2016kl} may be considered as a compromise between the two, because the posteriorgram probabilities for all possible phonemes in the source speech signals are estimated, converted and used to generate the target speech signals frame by frame.

Typically, text-independent VC requires parallel data. In other words, the data of utterance pairs produced by the source and target speakers for the same sentences are needed to train the conversion model.
But recently, methods based on deep learning using only non-parallel data have been proposed~\cite{hsu2017voice, chou2018multi, kaneko2017parallel}.
However, in these approaches due to the limitations of the conversion models or acoustic features used, the utterance length before and after conversion has to be kept the same, so the goal of reproducing the speaking rate and rhythmic patterns of the target speaker is simply impossible to realize.
Sequence-to-sequence learning performed on phoneme posteriorgram sequences may be a possible approach to achieve the above mentioned goal~\cite{miyoshi2017voice}, but all such approaches reported so far required parallel data.
In this paper, we propose a rhythm-flexible VC approach producing target speech signals of variable length but trained with non-parallel data only.

Below in~\autoref{sec:NPVC_DL}, we first introduce the related works on text-independent VC using deep learning trained with non-parallel data and the associated length constraint.
We then show in~\autoref{sec:VC_SS} a primarily text-dependent approach using sequence-to-sequence model to transform between source and target speakers over the phoneme posteriorgram sequences, which overcame the problem of length constraint but required parallel data.
In~\autoref{sec:proposed}, we therefore present the approach proposed here in this paper using non-parallel data but overcoming the length constraint to offer rhythm-flexible VC.
We list model architectures and implementation for this proposed approach in~\autoref{sec:model}, and show experimental results with evaluations in~\autoref{sec:experiments}.
Finally, we make some discussions and concluding remarks in ~\autoref{sec:dis_con}.

\section{Related Work}

\subsection{Non-parallel VC using Deep Learning}
\label{sec:NPVC_DL}
Recently, deep generative models such as Variational Autoencoders (VAEs)~\cite{kingma2013auto} and Generative Adversarial Networks (GANs)~\cite{goodfellow2014generative} including Conditional GANs (CGANs)~\cite{mirza2014conditional} were broadly studied because they can be applied to unsupervised learning problems. This is specially attractive for VC because that implies parallel data may not be needed.
With VAEs, the encoder first extracts a latent feature representing the speaker-independent linguistic content, and then the decoder is trained to generate the voice of the target speaker conditioned on the latent feature and some extra information regarding the target speaker~\cite{hsu2017voice, chou2018multi, hsu2016voice}.
With CGANs, with the guidance of the discriminator, the conditional generator tries to generate acoustic features sounding like being produced by the target speaker conditioned on the acoustic features produced by the source speaker.
Among the many extensions of CGANs, cycle-consistent adversarial network (Cycle-GAN)~\cite{zhu2017unpaired} and Star-GAN~\cite{choi2017stargan} have been very successfully used as domain translators between the source and target domains, and have been used for VC~\cite{kaneko2017parallel, gao2018voice, fang2018high, kameoka2018stargan}.

Although the above approaches are able to perform voice conversion without parallel data, the length of the generated signals are locked to be the same as that of the input signals due to the neural network architectures or the acoustic features used.
For example, some of them used combinations of recurrent neural networks (RNNs) and convolutional neural networks (CNNs)~\cite{hsu2017voice, chou2018multi, hsu2016voice} rather then the sequence-to-sequence encoder-decoder architecture.
These methods took a single frame or a segment of frames (e.g. 128 frames) as the input, and then generated a single frame or a segment of frames with the same length as the output.
Some other approaches chose Mel-cepstral coefficients (MCEPs), logarithmic fundamental frequency (log F0), and aperiodicities (APs) as the features, but the conversion was performed on MCEPs only~\cite{kaneko2017parallel, kameoka2018stargan}.
The converted MCEPs have to be of the same length as the original ones in order to be aligned with the sequences of log F0 and APs when synthesizing back to the waveform.
This fixed-length constraint makes it impossible for these very attractive deep learning approaches not requiring parallel data to be rhythm-flexible to better catch the prosody of the target speaker.

\subsection{Sequence-to-sequence Conversion over Posteriorgram Sequences Trained with Parallel Data}
\label{sec:VC_SS}
An approach utilizing Recurrent Neural Networks (RNNs) encoder-decoder for sequence-to-sequence learning~\cite{sutskever2014sequence} transforming the phoneme posteriorgram sequences between different speakers that can overcome the length constraint mentioned above was proposed~\cite{miyoshi2017voice}.
In this approach, in addition to a speech recognizer to produce the phoneme posteriorgram sequences and a speech synthesizer to reconstruct the signals, a module for transformation between the phoneme posteriorgram sequences for the source and target speakers was added in between to perform VC.
This latter transformation module includes an RNN encoder and an RNN decoder operating frame by frame.
The end-of-sequence token $\textless{EOS}\textgreater$ produced at the RNN decoder at any time removed the length constraint mentioned above and offered more flexible rhythm for the output speech.
However, the supervised training for sequence-to-sequence learning requires parallel data.
This leads to the new approach proposed in this paper, which offers rhythm-flexible VC with variable length but doesn't require parallel data, as is presented below.

\section{Proposed Approach}
\label{sec:proposed}
\begin{figure}[t]
  \centering
  \includegraphics[width=\linewidth]{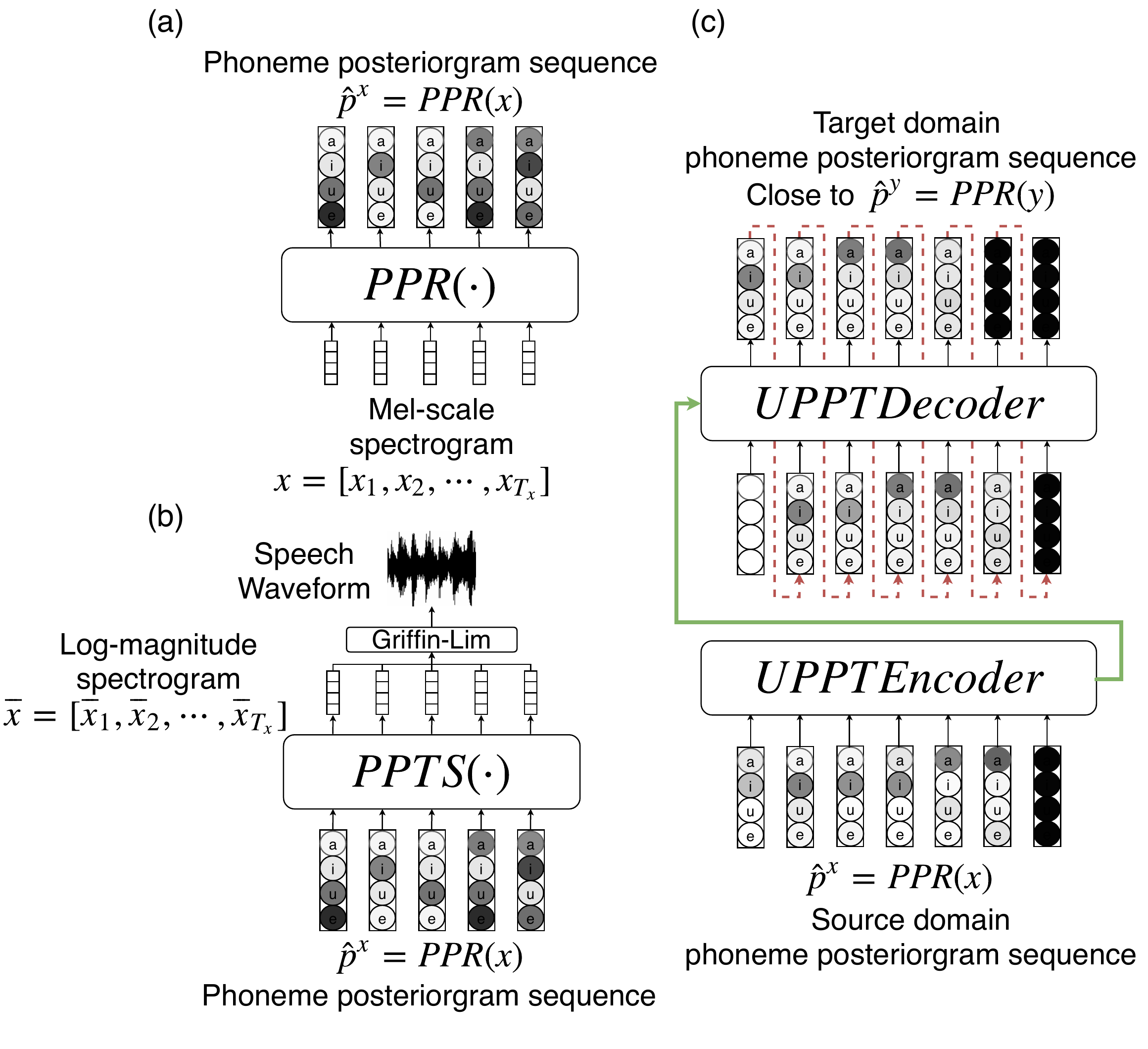}
  \caption{The three components of the proposed approach. $PPR(\cdot)$ in (a) stands for Phoneme Posteriorgram Recognizer and $PPTS(\cdot)$ in (b) for Phoneme-Posteriorgram-to-Speech Synthesizer. $UPPT$ in (c) stands for Unsupervised Phoneme Posteriorgram Transformer, which includes an encoder and a decoder. Dotted red arrows around the $UPPT$ decoder indicates the output at the previous time index is used as the input at the next time index. The green arrow indicates the final state of the encoder is fed to the initial state of the decoder.}
%
%
  \label{fig:model}
\end{figure}

The approach proposed here successfully overcomes the length constraint mentioned in~\autoref{sec:NPVC_DL} and removes the need for parallel data mentioned in~\autoref{sec:VC_SS} by adopting Cycle-GAN, which is an unsupervised style transfer model capable of transforming the phoneme posteriorgram sequences between speakers.
The three components of the approach is respectively presented in subsections~\ref{ssec:ppr},~\ref{ssec:ppts}, and~\ref{ssec:uppt} and~\autoref{fig:model} (a)(b)(c), the Cycle-GAN in~\autoref{ssec:cyclegan} and~\autoref{fig:cycle}, while the overview of the whole VC process is in~\autoref{fig:vc}.

\subsection{Phoneme Posteriorgram Recognizer}
\label{ssec:ppr}
As in~\autoref{fig:model} (a), let $x = [x_1, x_2, \cdots, x_{T_x}]$ and $y = [y_1, y_2, \cdots, y_{T_y}]$ be the acoustic feature vector sequences from the source and target speaker domains, $x_t$ and $y_t$ be the feature vector at time index $t$, and $T_x$ and $T_y$ be the lengths of $x$ and $y$. 
In~\autoref{fig:model} (a), $x$ and $y$ are the Mel-scale spectrogram.
Also, $l^x = [l^x_1, l^x_2, \cdots, l^x_{T_x}]$ and $l^y = [l^y_1, l^y_2, \cdots, l^y_{T_y}]$ are the ground truth label phoneme sequences corresponding to $x$ and $y$, respectively.
Phoneme Posteriorgram Recognizer $PPR(\cdot)$ is a speaker-independent neural network that estimates the phoneme posterior probabilities frame by frame given an acoustic feature vector sequence. 
This recognizer $PPR(\cdot)$ is trained to minimize $L_{xent}(l^{x}, PPR(x))$ and $L_{xent}(l^{y}, PPR(y))$, which are the cross-entropy between the ground truth label sequences (an one-hot vector for each time $t$) and the estimated phoneme posteriorgram sequences for data in both source and target speaker domains.

\subsection{Phoneme-Posteriorgram-to-Speech Synthesizer}
\label{ssec:ppts}
As in~\autoref{fig:model} (b), Phoneme-Posteriorgram-to-Speech Synthesizers $PPTS_x(\cdot)$ and $PPTS_y(\cdot)$ are the reverse process of $PPR(\cdot)$ for data in the source and target speaker domains respectively, or two neural networks that predict the speech feature vectors $\Bar{x}$ and $\Bar{y}$ frame by frame given the phoneme posteriorgram sequences $\hat{p}^x = PPR(x)$ and $\hat{p}^y = PPR(y)$.
In~\autoref{fig:model} (b), $\Bar{x}$ and $\Bar{y}$ are the log-magnitude version of $x$ and $y$. Griffin-Lim is the algorithm synthesizing the speech waveform from the predicted log-magnitude version $\Bar{x}$ and $\Bar{y}$~\cite{griffin1984signal}.
$PPTS_x(\cdot)$ and $PPTS_y(\cdot)$, are respectively trained to minimize the mean squared error between the ground truth speech feature vectors and the reconstructed version, $L_{mse}(\Bar{x}, PPTS_x(\hat{p}^x))$ and $L_{mse}(\Bar{y}, PPTS_y(\hat{p}^y))$.

\subsection{Unsupervised Phoneme Posteriorgram Transformer}
\label{ssec:uppt}
As shown in~\autoref{fig:model} (c), the Unsupervised Phoneme Posteriorgram Transformer $UPPT$ is an attention-based sequence-to-sequence model including an $UPPT$ encdoer and an $UPPT$ decoder, which transforms a source domain posteriorgram sequence $\hat{p}^x = PPR(x)$ into another posteriorgram sequence very close to those for signals in the target domain, $\hat{p}^y = PPR(y)$.
The green arrow indicates the final state of the encoder is fed to the initial state of the decoder, and the dotted red arrows around the $UPPT$ decoder indicate the output of the previous time index is used as the input at the next time index.
This is a sequence-to-sequence model used to remove the length constraint and achieve the rhythm-flexible VC mentioned previously.

\subsection{Cycle-GAN}
\label{ssec:cyclegan}
\begin{figure}[t]
  \centering
  \includegraphics[width=\linewidth]{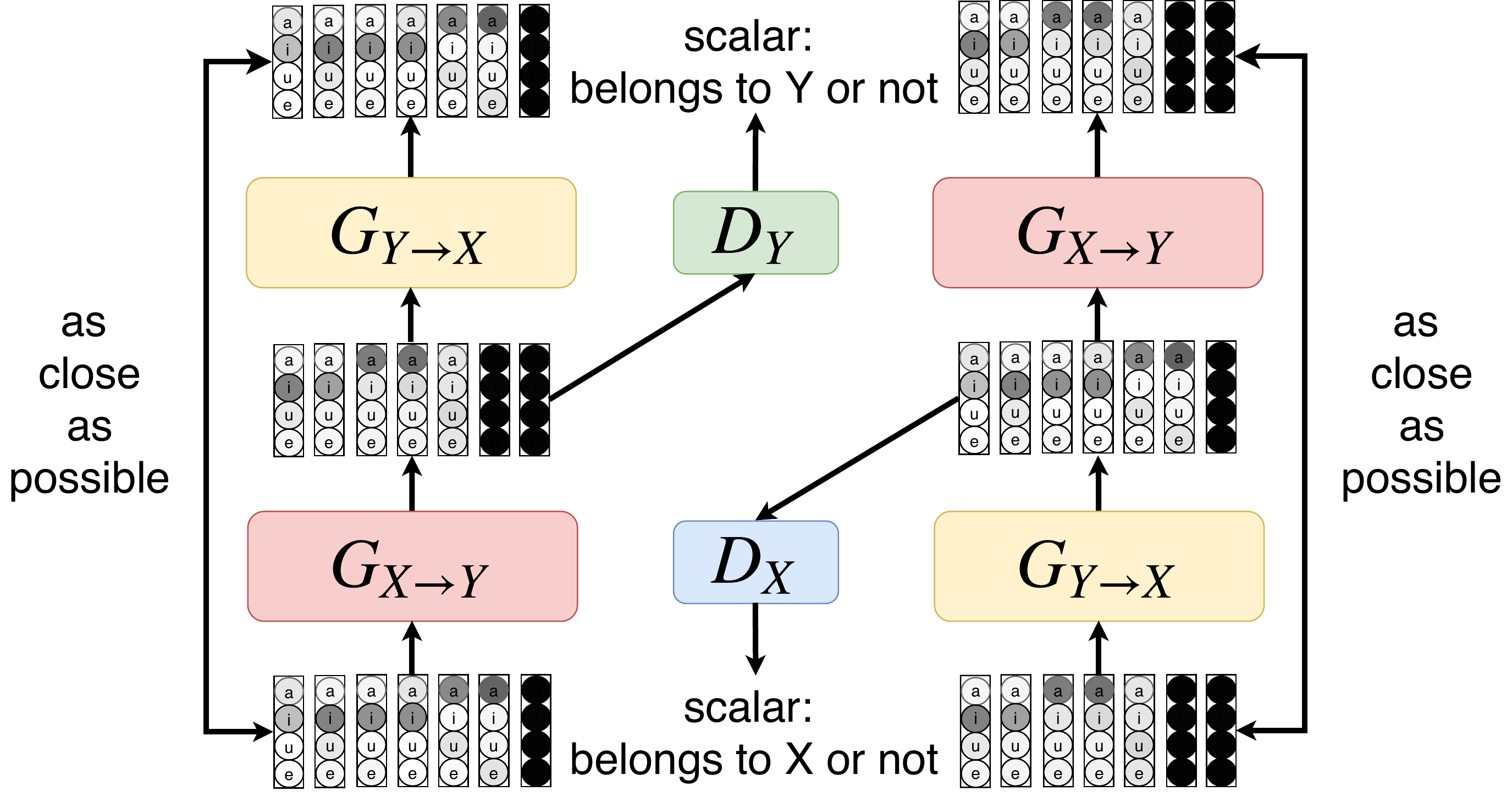}
  \caption{Cycle-GAN. $G_{X\rightarrow{Y}}$ and $G_{Y\rightarrow{X}}$ refer to generators. $D_{X}$ and $D_{Y}$ refer to discriminators. Blocks with the same color share the same set of neural network parameters. Each generator is built with a pair of $UPPT$ encoder and $UPPT$ decoder.
  }
  \label{fig:cycle}
\end{figure}

Let $X$ and $Y$ be the two sets that contain all estimated phoneme posteriorgram sequences $\hat{p}^x$, $\hat{p}^y$ from the source and target speaker domains respectively. 
We adopt here the cycle-consistent generative adversarial network (Cycle-GAN) to learn the mapping between $X$ and $Y$ without paired data.
As shown in~\autoref{fig:cycle}, the whole training procedure includes two sets of generative adversarial networks (GANs), each with a generator and a discriminator.
After Cycle-GAN training, two generators, $G_{X\rightarrow{Y}}$ and $G_{Y\rightarrow{X}}$ are obtained.
These two generators are two transformers ($UPPT$s in~\autoref{ssec:uppt}) that transform $\hat{p}^x$ to $\hat{p}^{x\rightarrow{y}}$ ($\hat{p}^{x\rightarrow{y}}=G_{X\rightarrow{Y}}(\hat{p}^x)$) and $\hat{p}^y$ to $\hat{p}^{y\rightarrow{x}}$ ($\hat{p}^{y\rightarrow{x}}=G_{Y\rightarrow{X}}(\hat{p}^y)$) respectively, where $\hat{p}^{x\rightarrow{y}}$ is a phoneme posteriorgram sequence mapped from the source domain to target domain and $\hat{p}^{y\rightarrow{x}}$ vice versa.
Two discriminators are also trained, $D_X$　and $D_Y$, to discriminate whether a phoneme posteriorgram sequence is a real one generated from a signal in a domain, or a fake one transformed from another domain.

\subsubsection{Training Goal of Generators (or UPPTs)}
\label{sssec:gen}
The generators $G_{X\rightarrow{Y}}$, $G_{Y\rightarrow{X}}$ take the phoneme posteriorgram sequences from a speaker domain as the input and produce another phoneme posteriorgram sequence close to those for another speaker domain.
Both are built with attention-based sequence-to-sequence model to learn the mapping between $\hat{p}^x \in X$ and $\hat{p}^y \in Y$ such that the distribution of $G_{X\rightarrow{Y}}(\hat{p}^x)$ is as indistinguishable from that of $\hat{p}^y$ as possible, and $G_{Y\rightarrow{X}}(\hat{p}^y)$ is as indistinguishable from $\hat{p}^x$ as possible.
The training of the generators are guided by the discriminators described below to achieve the above scenario.

\subsubsection{Training Goal of Discriminators}
\label{sssec:dis}
The discriminators $D_X$ and $D_Y$ take a phoneme posteriorgram sequence as the input, and produce a scalar indicating how "real" the input is from the sets $X$ or $Y$ for the domain considered, or actually fake or transformed from another domain.
Such discriminators are to guide the generators.
So the training objective of discriminators are to distinguish between the real sequences such as $\hat{p}^x$, $\hat{p}^y$ and fake sequences such as $G_{Y\rightarrow{X}}(\hat{p}^y)$, $G_{X\rightarrow{Y}}(\hat{p}^x)$ generated by the generators, and give higher scores to real ones and lower scores to fake ones.

\subsubsection{Objective functions}
\label{sssec:obj}
Several objective functions are defined here as given below.
1. Adversarial Loss: 
Adversarial losses~\cite{goodfellow2014generative} are applied to both mapping functions, $G_{X\rightarrow{Y}}$ and  $G_{Y\rightarrow{X}}$. 
For the mapping function $G_{X\rightarrow{Y}}$ and its discriminator $D_Y$, we express the objective as in~\eqref{eq1}. 
\begin{equation}
    \scalebox{0.92}{
    \begin{math}
    \begin{aligned}
    L_{GAN}(G_{X\rightarrow{Y}}, D_Y) =
    & \ E_{{y}\sim{Y}}[{\log{D_Y(y)}}]\\
    + & \ E_{{x}\sim{X}}[{\log{(1-D_Y(G_{X\rightarrow{Y}}(x)))}}].
    \end{aligned}
    \end{math}
    }
  \label{eq1}
\end{equation}
$L_{GAN}(G_{Y\rightarrow{X}}, D_X)$ is defined in exactly the same way as~\eqref{eq1}, except the roles of $X$ and $Y$ are reversed.

2. Cycle Consistency Loss:
Cycle consistency losses~\cite{zhu2017unpaired} are applied when training the two generators.
The transform cycle should be able to bring $x$ back to the original phoneme posteriorgram sequence, i.e. $G_{Y\rightarrow{X}}(G_{X\rightarrow{Y}}(x)) \approx x$ and $G_{X\rightarrow{Y}}(G_{Y\rightarrow{X}}(y)) \approx y$. We express this objective as in~\eqref{eq2}. (Note that $L_{xent}$ in~\eqref{eq2} means cross-entropy)
\begin{equation}
    \scalebox{0.92}{
    \begin{math}
    \begin{aligned}
    L_{cycle}(G_{X\rightarrow{Y}}, G_{Y\rightarrow{X}})=
    & \ E_{{x}\sim{X}}[L_{xent}(x, {G_{Y\rightarrow{X}}(G_{X\rightarrow{Y}}(x))})]\\
    + & \ E_{{y}\sim{Y}}[L_{xent}(y, {G_{X\rightarrow{Y}}(G_{Y\rightarrow{X}}(y))})].
    \end{aligned}
    \end{math}
    }
  \label{eq2}
\end{equation}

3. Identity Mapping Loss:
Identity mapping loss as proposed in the original work of Cycle-GAN~\cite{zhu2017unpaired} is also used here.
When real samples of the target domain are provided as the input to the generator, the transformed result should be as close to the input as possible.
It was found that adding this objective as an extra regularization term for the generators actually improved the transformed results.
We express this objective as in~\eqref{eq3}.
\begin{equation}
    \scalebox{0.92}{
    \begin{math}
    \begin{aligned}
    L_{identity}(G_{X\rightarrow{Y}}, G_{Y\rightarrow{X}}) =
    & \ E_{{x}\sim{X}}[L_{xent}(x, {G_{Y\rightarrow{X}}(x))})]\\
    + & \ E_{{y}\sim{Y}}[L_{xent}(y, {G_{X\rightarrow{Y}}(y))})].
    \end{aligned}
    \end{math}
    }
  \label{eq3}
\end{equation}

The full objective for Cycle-GAN training is the sum of (1)(2)(3):
\begin{equation}
    \scalebox{0.92}{
    \begin{math}
    \begin{aligned}
        L_{cycle\_gan}(G_{X\rightarrow{Y}}, G_{Y\rightarrow{X}}, D_X, D_Y) =\\
        L_{GAN}(G_{X\rightarrow{Y}}, D_Y) + L_{GAN}(G_{Y\rightarrow{X}}, D_X)\\
        +\lambda_1{L_{cycle}(G_{X\rightarrow{Y}}, G_{Y\rightarrow{X}})}+\lambda_2{L_{identity}(G_{X\rightarrow{Y}}, G_{Y\rightarrow{X}})},\\
    \end{aligned}
    \end{math}
    }
  \label{eq4}
\end{equation}
where $\lambda_1, \lambda_2$ are balancing parameters. So overall we aim to solve:
\begin{equation}
    \scalebox{0.92}{
    \begin{math}
    \begin{aligned}
        G_{X\rightarrow{Y}}^*, G_{Y\rightarrow{X}}^* = \arg\min_{G_{X\rightarrow{Y}}, G_{Y\rightarrow{X}}}\max_{D_X,D_Y}(L_{cycle\_gan}).
    \end{aligned}
    \end{math}
    }
  \label{eq5}
\end{equation}

\begin{figure}[t]
  \centering
  \includegraphics[scale=0.42]{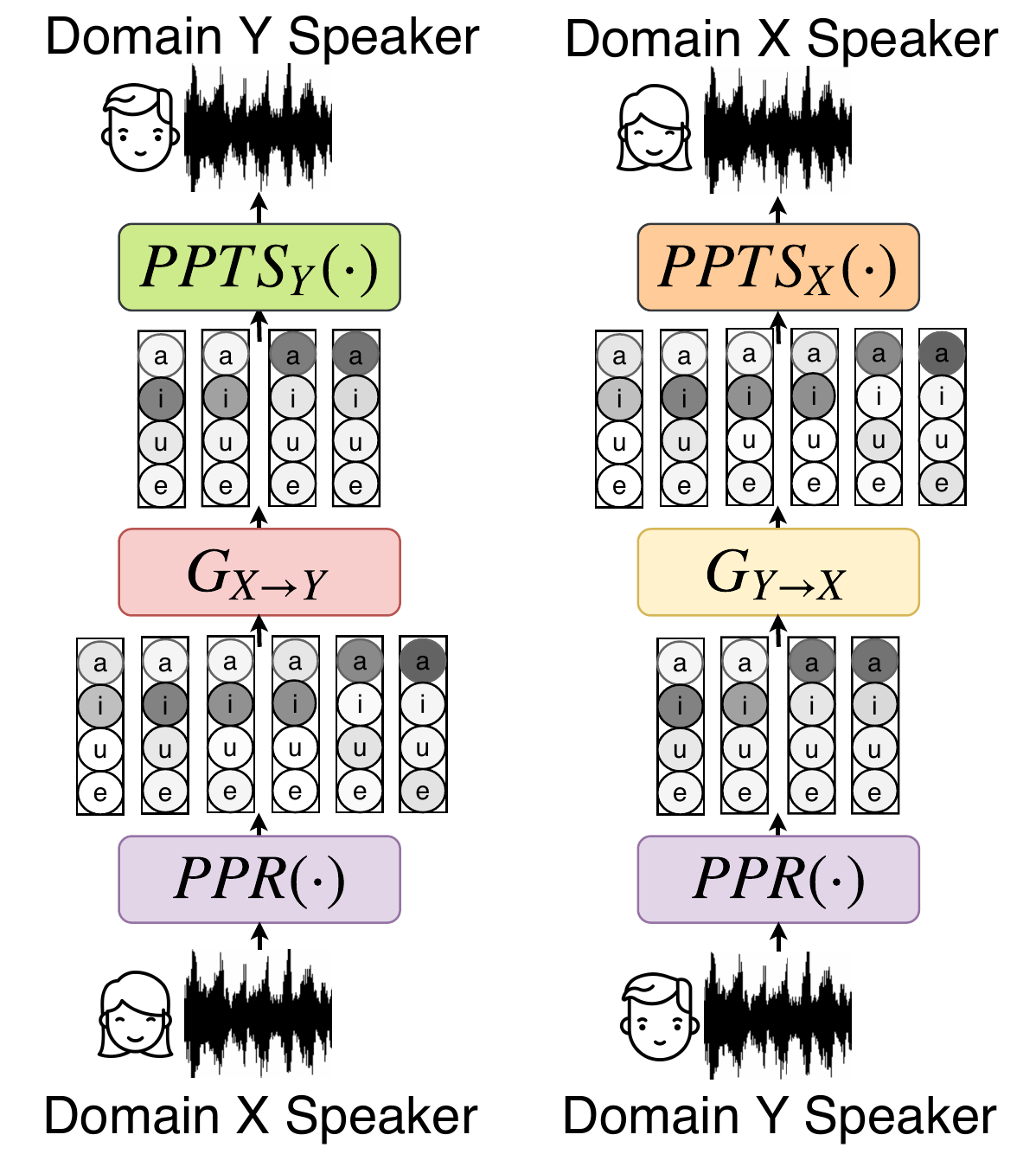}
  \caption{The complete voice conversion process for the proposed approach. Blocks with the same color share the same set of neural network parameters.}
  \label{fig:vc}
\end{figure}

\subsection{Overall Voice Conversion}
\label{ssec:overall_vc}
As shown in~\autoref{fig:vc}, the overall voice conversion is achieved by first passing the source speaker's speech signal through $PPR(\cdot)$ in~\autoref{ssec:ppr} to obtain its phoneme posteriorgram sequence, then using $UPPT$ in~\autoref{ssec:uppt} to transform it to the target domain, and finally using $PPTS(\cdot)$ in~\autoref{ssec:ppts} trained on the target domain to synthesize the target speech signal from the given phoneme posteriorgram sequence.

\section{Model Implementation and Training}
\label{sec:model}
We adopted primarily the model architecture from the CBHG module~\cite{lee2016fully} for all three parts of the proposed approach including the phoneme posteriorgram recognizer $PPR(\cdot)$, phoneme-posteriorgram-to-speech synthesizer $PPTS(\cdot)$ and the unsupervised phoneme posteriorgram transformer $UPPT$ (or the generators).
The convolution-bank in CBHG module was found to be able to better capture the local information over time, reduce overfitting and generalize well to long and complex inputs, such as acoustic feature sequences~\cite{wang2017tacotron}.
As previously suggested~\cite{kaneko2017parallel}, for the discriminators, we treated the input phoneme posteriorgram sequences as pictures with channel size one, and performed several 1D-convolution layers with strides larger then one for better capturing the local properties such as how many frames a speaker usually needs to produce a specific phoneme.
The attention mechanism in $UPPT$ was shown to be able to effectively improve the decoder's prediction~\cite{bahdanau2014neural}.
The overall model architecture and training details are available\footnote{ https://github.com/acetylSv/rhythmic-flexible-vc-arch} but left out here for space limitation. We considered $PPR(\cdot)$ as a pseudo-labeler. We first trained $PPR(\cdot)$ with the objective mentioned in~\autoref{ssec:ppr}, and then trained $PPTS(\cdot)$ with the objective mentioned in~\autoref{ssec:ppts}.
With the above done, we then collected the estimated results of $PPR(\cdot)$ to train $UPPT$.

For $PPR(\cdot)$, we used mel-scale spectrogram as the input acoustic features, and the phoneme set defined in Carnegie Mellon pronouncing dictionary~\cite{weide2005carnegie} as the labels for the posteriorgram sequences.
Thus the input to $UPPT$ were sequences of vectors with dimension 70 (39 phoneme types with stress combinations, each treated as mono-phoneme).
For $PPTS$, we used log-magnitude spectrogram as the output acoustic features, over which Griffin-Lim algorithm~\cite{griffin1984signal} was applied to synthesize the waveform.
All other detailed setting followed the previous work~\cite{wang2017tacotron}.

\begin{figure}[t]
  \centering
  \includegraphics[scale=0.48]{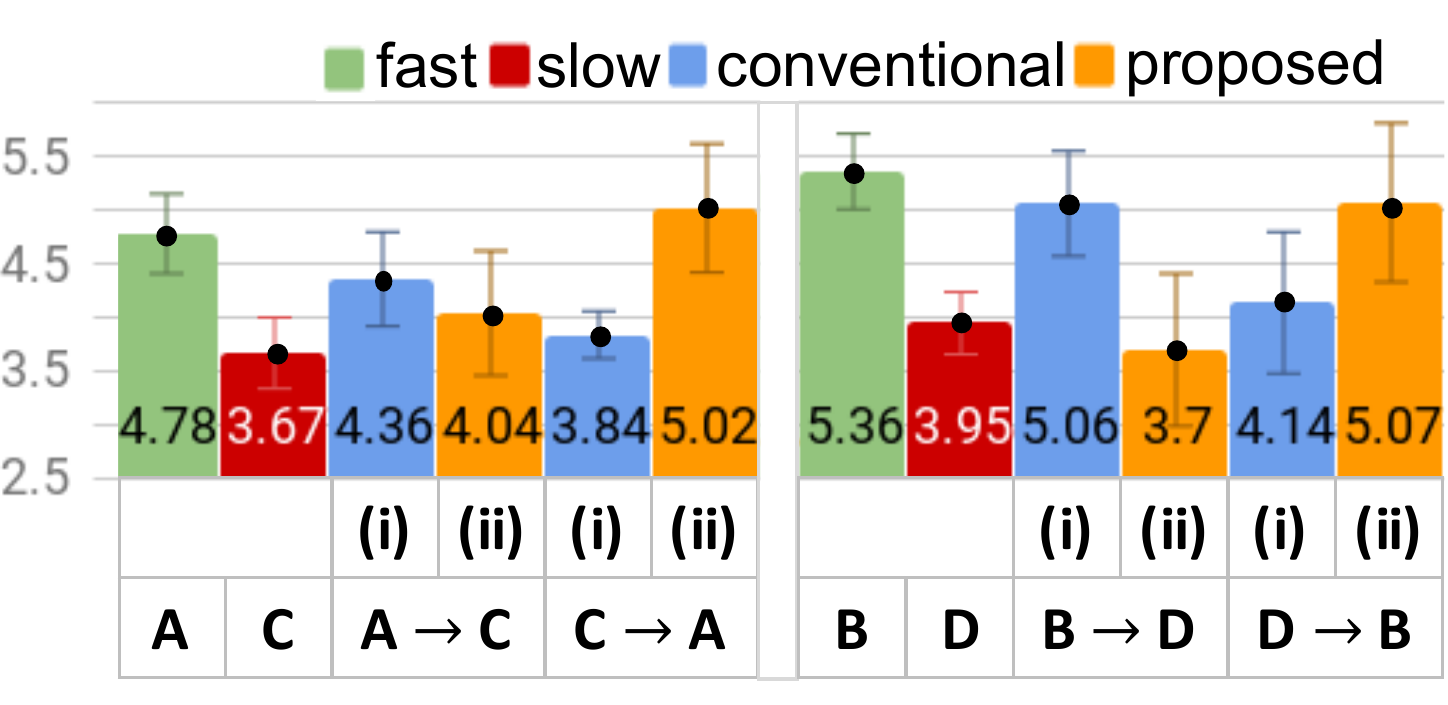}
  \caption{Average speaking rates (number of syllables / sec) for utterances in testing set before and after conversion. The dots and bars indicate the averages and the standard deviations. Speakers A, B belonged to the fastest speaking group and C, D to the slowest speaking group, (i) achieved by the conventional method~\cite{sun2016phonetic} while (ii) by the proposed approach. The numbers shown are the averages.}
  \label{fig:sr}
\end{figure}

\begin{figure*}[t]
  \centering
  \includegraphics[scale=0.46]{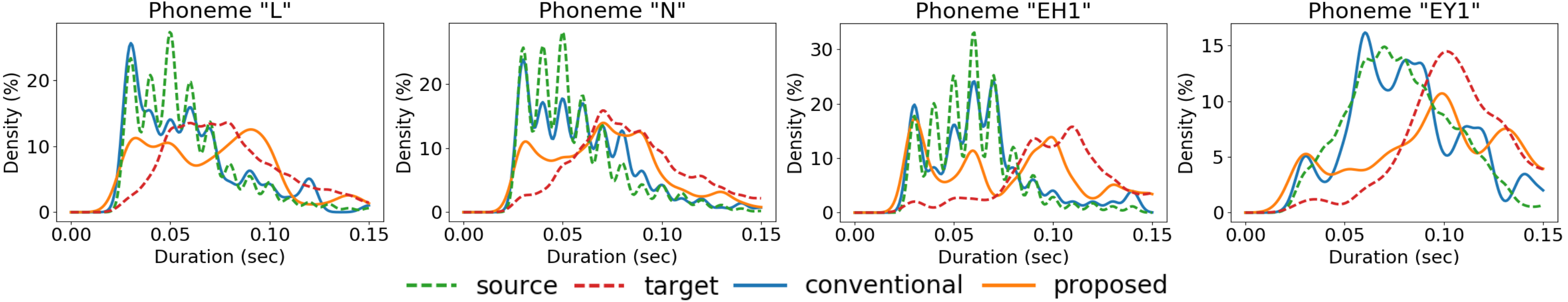}
  \caption{Example rhythmic patterns (duration distributions) for phonemes "L", "N", "EH1", "EY1". The histograms were normalized by Gaussian kernel density estimation (bandwidth=0.125). Different colors represent the rhythmic patterns for different speakers (green and red) and converted voice (blue and brown).}
  \label{fig:pattern}
\end{figure*}

\section{Experiments and Results}
\label{sec:experiments}
\subsection{Experimental setup}
\label{ssec:exp_setup}
We used two datasets under a fully non-parallel setting.
One is Librispeech~\cite{panayotov2015librispeech}, an audio book read by multi-speakers. 
The other one is VCTK~\cite{veaux2017cstr}, which is a multi-speaker dataset primarily reading newspapers and elicitation paragraphs intended to identify the speaker's accent.
Both datasets were randomly split into training, validation and testing sets with percentages of 80\%, 10\% and 10\%. 
The phone boundaries were not available in both datasets, so we used a force-aligner pretrained on Librispeech dataset~\cite{mcauliffe2017montreal} to get the phone boundaries and corresponding phone classes for training $PPR(\cdot)$.
\subsubsection{Librispeech Dataset}
\label{sssec:libri_setup}
Using Praat Script Syllable Nuclei~\cite{de2009praat} to measure the speaking rate, we picked the fastest 20 speakers and the slowest 20 speakers to form a subset with a total length of 15.8 hours or 4609 utterances for evaluation of conversion across different speaking rates.
When training the three components in~\autoref{fig:model}, we used all the 40 speakers to train the speaker-independent $PPR(\cdot)$, the grouped fastest and slowest 20 speakers as two domains to initialize the Cycle-GAN training  for $UPPT$ (followed by individual training for each conversion pair), and individually trained speaker-dependent $PPTS(\cdot)$ for each speaker.
\subsubsection{VCTK Dataset}
\label{sssec:vctk_setup}
We chose 18 speakers, some native and some non-native, with a total of 7.3 hours or 7132 utterances, as a different scenario for rhythmic patterns.
We used all the 18 speakers to train the speaker-independent $PPR(\cdot)$, but trained the $UPPT$ and $PPTS(\cdot)$ for each conversion pair individually.

\subsection{Objective Evaluation}
\label{ssec:obj}
To show the proposed approach is able to learn the speaking rates of the target speakers, we chose two speakers A, B (with IDs 6925 and 460) from the fastest speaking group of Librispeech and two speakers C, D (with IDs 163 and 1363) from the slowest speaking group and performed the conversions A $\leftrightarrow$ C and B $\leftrightarrow$ D on the utterances in their testing sets using a conventional method~\cite{sun2016phonetic} and the proposed approach.
This is actually an ablation study, since the only difference between the two is whether the $UPPT$ proposed here was used or not.
The results are plotted in~\autoref{fig:sr}, where the averages and standard deviations of the speaking rates are shown for the two approaches.
We can see from~\autoref{fig:sr} the proposed approach could mimic the speaking rates of the target speaker much better.

To show the proposed approach is capable of learning the rhythmic patterns (phoneme duration distributions) for the target speaker, we chose a pair of speakers E, F (with IDs p231 and p265) from VCTK and performed the conversion E $\rightarrow$ F on their testing utterances.
We used the pretrained force-aligner to obtain the phoneme duration and normalized the histograms by Gaussian kernel density estimation.
The example rhythmic patterns for two vowels and two consonants are plotted in~\autoref{fig:pattern}, in which the different colors are respectively for source and target speakers (green and red) and the converted voice by a conventional~\cite{sun2016phonetic} (blue) and the proposed (brown) approaches.
We can see from~\autoref{fig:pattern} different speakers did show very different rhythmic patterns, and the proposed approach was able to mimic these patterns of the target speaker much better.

\subsection{Subjective Evaluation}
\label{ssec:sub}
\begin{figure}[t]
  \centering
  \includegraphics[scale=0.40]{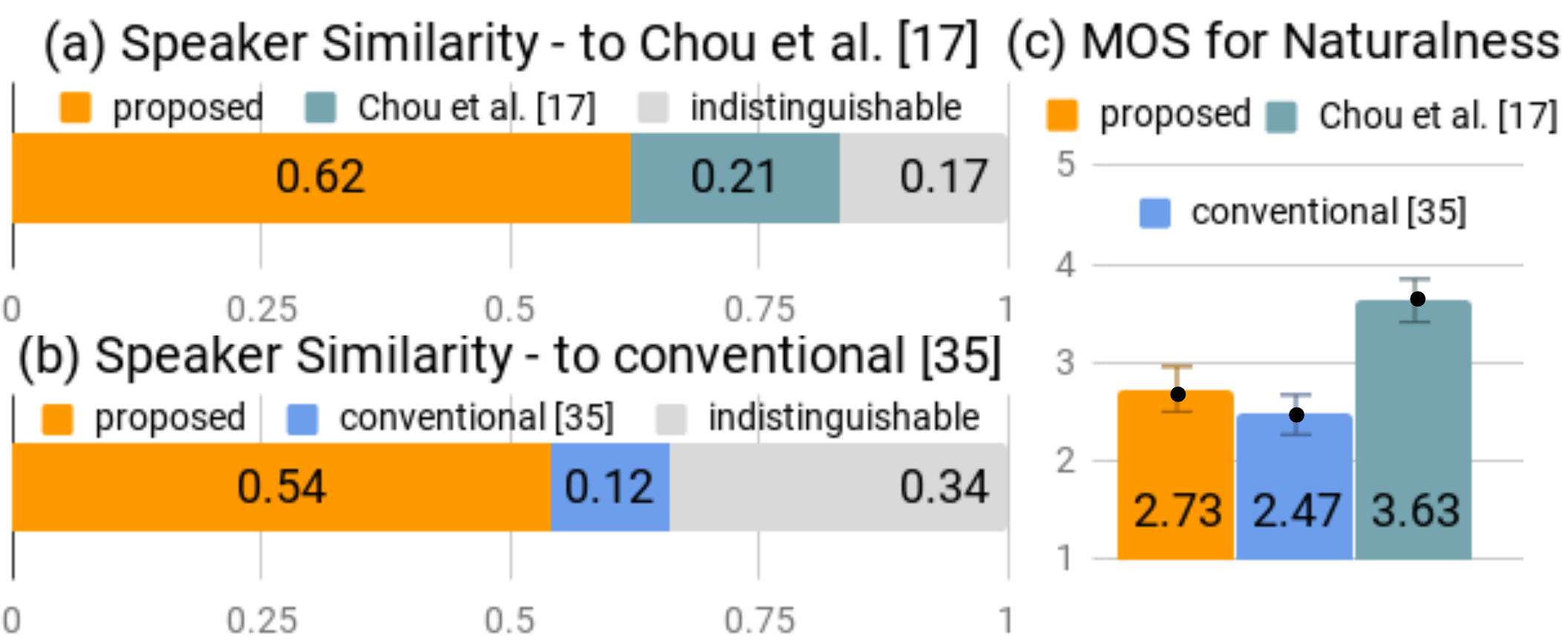}
  \caption{Subjective evaluation results: binary preference test for speaker similarity compared to (a) the recently proposed non-parallel VC with length constraint (Chou et al.~\cite{chou2018multi}), (b) the conventional method~\cite{sun2016phonetic} (ablation study); and (c) 5-scale naturalness MOS scores similarly compared.}
  \label{fig:sub_res}
\end{figure}

Subjective evaluation was performed on converted voice (including both intra-gender and inter-gender conversions) from VCTK datasets.
In the binary preference test for speaker similarity, 20 subjects were given pairs of converted voice in random order and asked to choose one sounding more similar to a reference target utterance produced by the target speaker, comparing the proposed approach to a recently proposed non-parallel VC by Chou et al.~\cite{chou2018multi} and the conventional method~\cite{sun2016phonetic}.
The results are in~\autoref{fig:sub_res} (a)(b).
We can see the proposed approach obviously outperformed the two previous approaches in terms of speaker similarity.

The MOS for naturalness in~\autoref{fig:sub_res} (c) shows the proposed approach is better than the conventional method~\cite{sun2016phonetic}, although not as good as the recently proposed non-parallel VC~\cite{chou2018multi}, very probably because of the mean square error (MSE) objective function used in training $PPTS(\cdot)$ in~\autoref{ssec:ppts}.
It was found that models trained with MSE objective tend to output average predictions~\cite{kaneko2017generative}, which may lead to over-smoothed log-magnitude spectrograms and blurred sounds after the Griffin-Lim algorithm.
Investigations for replacing Griffin-Lim vocoder with a neural vocoder~\cite{van2016wavenet} or applying post-filters to enhance the output log-magnitude spectrograms are under progress.
Another possible direction may be applying sequence-to-sequence Cycle-GAN directly on log-magnitude spectrograms rather than on the phoneme posteriorgram sequences, but at the difficulties of the high feature dimension and complex model structures.

\section{Conclusion}
\label{sec:dis_con}
Objective and subjective evaluation on two different datasets showed that the proposed approach is able to mimic the voice characteristics of a target speaker, including the speaking rate and rhythmic patterns, without parallel data by utilizing sequence-to-sequence learning trained with Cycle-GAN to remove the length constraint.
Although phoneme boundaries are needed for the training data, an easily obtained pretrained force-aligner can offer these boundaries.

\end{document}